\documentclass[a4paper,12pt]{article}
\usepackage{amsmath}
\usepackage{citehack}
\usepackage{bm}
\usepackage{graphicx}
\usepackage{epstopdf}
\usepackage{color}
\usepackage{comment}
\usepackage{hyperref}
\usepackage{float}
\usepackage{tikz, xcolor, pgfplots}
\usetikzlibrary{shapes, arrows}
\usepackage{multirow}

\begin{document}

\title{Molecular dynamic simulation of wet water vapour transport in porous medium
\thanks{topic JINR LIT No. 05-6-1118-2014/2019, protocol No. 4596-6-17/19. HybriLIT resources.}}
\author{E.G. Nikonov$^1$,  M. Popovi\v{c}ov\'a$^2$\\
\begin{minipage}{10cm}
\begin{center}
\small\em
\vspace{4mm}
$^1$Joint Institute for Nuclear Research,\\ 141980 Dubna, Moscow Region, Russia\\ email: e.nikonov@jinr.ru \\
\vspace{4mm}
$^2$University of Pre\v{s}ov,\\ str. Kon\v{s}tantinova 16, 080 01 Pre\v{s}ov,  Slovakia\\ email: maria.popovicova@unipo.sk
\end{center}
\end{minipage}
} 
\date{}
\maketitle 

\begin{abstract}
Studies of the transport of wet water vapour are relevant for various areas of human activity, including the construction and production of building materials, mining, agriculture, environmental safety of technological processes, scientific research. In particular, one of the methods for extracting highly viscous bitumen grades of oil from the subterranean depths is based on the dilution of the contents of the porous medium by means of pumping coolants. The simplest and environmentally friendly coolant is wet steam containing small drops of water. In addition, with the help of heated water vapour, the filtering elements of collectors are cleaned from sediments of the solid phase (for example, paraffins, gas hydrates and ice floes) on the walls of the porous medium.
For realistic modeling of the processes of filtration and heat and mass transfer during the injection of wet steam into a porous medium, it is necessary to investigate the characteristics of the interaction of saturated water vapour with individual through-type pores. In this paper, a study was carried out through mathematical modelling of the dependence of the diffusion rate of wet water vapour on the pressure difference outside the pore that occurs when water vapour is injected into a porous medium. The dependencies of the diffusion rate on the pore cross section, the magnitude of vapour adsorption on the pore walls, as well as the effect of water vapour temperature on all these processes were also investigated. Practical interest is the study of the influence of the rate of cooling of water vapour on the diffusion rate and the adsorption of water vapour on the wall of the pores. The calculations were carried out using a hybrid type model that combines molecular dynamics and macro-diffusion approaches to describe the interaction of water vapour with individual pores.

\end{abstract}

\section{Introduction}

Investigations of wet water vapour transport phenomena are very actual for various areas of human activity, including the construction and production of building materials \cite{Krus1996}, mining, agriculture, environmental safety of technological processes, scientific research. In particular, one of the methods for extracting highly viscous bitumen grades of oil from the subterranean depths is based on the dilution of the contents of the porous medium by means of pumping coolants. The simplest and environmentally friendly coolant is wet steam containing small drops of water \cite{Shagapov2004}. In addition, with the help of heated water vapour, the filtering elements of collectors are cleaned from sediments of the solid phase (for example, paraffins, gas hydrates and ice floes) on  walls of the porous medium.

One of the most interesting phenomenon for scientific research in various fields of engineering as, e.g., building physics and earth sciences is  an absorption of moisture in unsaturated porous media \cite{Janetti2018}. It has been shown \cite{Janetti2018} that, under certain assumptions, the moisture transfer obeys a diffusion law in which the local moisture content is the driving potential, while the diffusivity is a function of the moisture content itself. This diffusion
law is suitable to predict the transient moisture absorption in homogeneous, capillary active, construction materials and soils if gravity is negligible when compared to the capillary forces. Based on the foregoing, for a correct description of porous materials behavior during wetting and drying, it is necessary to know with a sufficient degree of accuracy the dependence of the diffusivity on moisture content in the porous material at given time moment.

For realistic modelling of wet steam penetration in and out porous medium, it is necessary to investigate the characteristics of the interaction of saturated water vapour with individual through-type pores. It could be useful for understanding of the same processes in the bulk porous media with close to uniform distribution of pores or capillaries. In this paper, studies of the dependence of the diffusivity on the moisture content of a porous material are carried out by using molecular dynamic simulations. All calculations have been made for the case of two-dimensionl model of individual through-type pores.  There was also investigated a dependence of the diffusion rate of wet water vapour on the pressure difference outside the pore that occurs when water vapour is injected into porous medium or when the porous material simply  dries or gets wet. Dependences of the diffusion rate on the pore cross section, the magnitude of water vapour adsorption on the pore walls, as well as the effect of water vapour temperature on all these processes were also researched. Practical interest is the study of the influence of the rate of cooling of water vapour on the diffusion rate and the adsorption of water vapour on the wall of the pores. Calculations were carried out using software for molecular dynamic simulations developed in \cite{NPP1709, NPP1708}.

\section{General problem}
In accordance with the general approach to the description of diffusion processes water vapour passage through a porous media can be done by diffusion equation in the following form.
\begin{equation}
\frac{\partial w(\mathbf{r},t)}{\partial t}=\nabla\cdot\left[ D(w,\mathbf{r}) \nabla w(\mathbf{r},t)\right], 
\end{equation}
where $w(\mathbf{r},t)$ is the density of the water vapour (or humidity) at location $\mathbf{r}$ and time $t$ and $D(w,\mathbf{r})$ is the generalized diffusion coefficient (or diffusivity) for density $w$ at location $\mathbf{r}$. If the diffusion coefficient depends on the water vapour density then the equation is nonlinear, otherwise it is linear.

More generally, when $D$ is a symmetric positive definite matrix, the equation describes anisotropic diffusion, which is written (for three dimensional diffusion) as:
\begin{equation}
\frac{\partial w(\mathbf{r},t)}{\partial t}=\sum_{i=1}^{3}\sum_{j=1}^{3}\frac{\partial}{\partial x_i}\left[D_{ij}(w,\mathbf{r})\frac{\partial w(\mathbf{r},t)}{\partial x_j}\right].
\end{equation}

If $D$ is constant, then the equation reduces to the following linear differential equation:
\begin{equation}
\frac{\partial w(\mathbf{r},t)}{\partial t}= D \nabla^2 w(\mathbf{r},t). 
\end{equation}

In the case of isothermal absorption in a material which can be assumed homogeneous at the
macroscopic scale and where gravity is negligible when compared to the capillary forces the diffusion of water vapour through the porous media can be described by one-dimensional moisture transfer model with the following diffusion equation \cite{Janetti2018}.
\begin{equation}
\frac{\partial w(x,t)}{\partial t}= \frac{\partial}{\partial x}\left(D(w)\frac{\partial w(x,t)}{\partial x}\right).
\label{Dif1dim} 
\end{equation}
Here $w(x,t)$ denotes the local volumetric water content $(m^3/m^3)$ and $D(w)$ the diffusivity
$(m^2/s)$, while the initial and boundary conditions are given by the following equations:
\begin{eqnarray*}
w(0,x)&=&w_0<1\\
w(t,0)&=&1\\
\lim_{x\rightarrow\infty} w(t,x) &=& w_0 
\end{eqnarray*}
To solve the problem given by the equation  (\ref{Dif1dim}), the diffusivity $D(w)$, which characterizes the
material behavior, has to be known. Experimental data show that this function presents generally a nearly
exponential growth. In fact, approximated solutions for an exponential diffusivity curve have been already published in various studies \cite{Janetti2018}. In all approaches currently existing in the case of the dependence of the diffusivity  on the local volumetric water content in a porous material, direct methods are used to solve the diffusion equation. These can be numerical methods based on the finite difference or finite element methods. This may be a method of approximating the original nonlinear problem and solving it either analytically or numerically under the assumption that the solution of this approximating problem in some sense tends to the solution of the original problem. One of the way to solve the problem  (\ref{Dif1dim}) analytically is the change of the original diffusivity function by multiple step function which is equal a constant in each step interval \cite{Janetti2018}. This approach proved to be very effective for solving problem for the one-dimensional diffusion model (\ref{Dif1dim}). 

\begin{figure}[H]
\begin{center}
\includegraphics[width=0.6\linewidth]{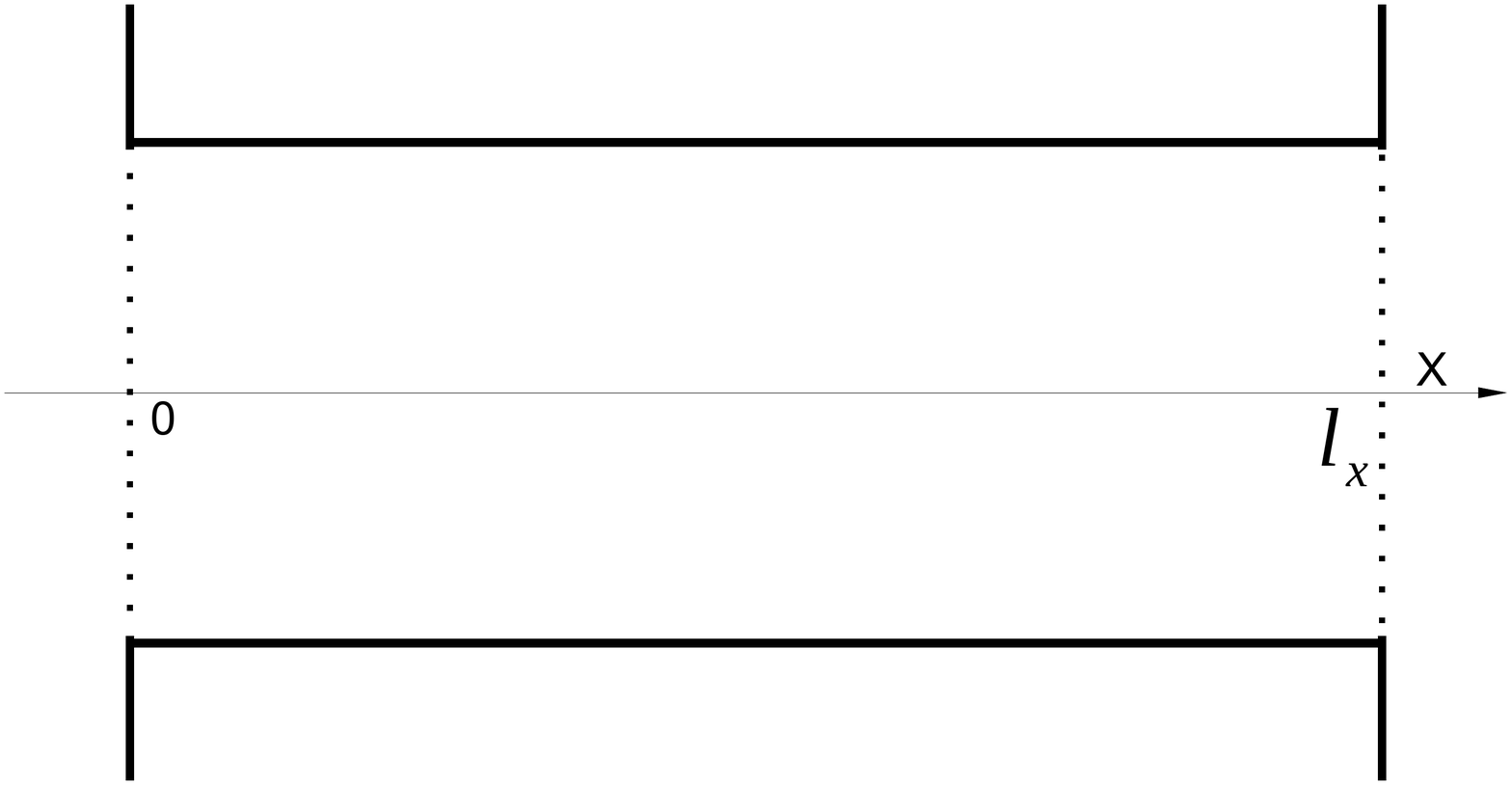}
\end{center}
\vspace{-1cm}
\caption{The shape of two-dimensional through type pore}
\label{1D-pore_diff}
\end{figure}

In this paper it is proposed a method for solving problem (\ref{Dif1dim}) based on the use of molecular dynamic simulation to obtain the dependence of the diffusion coefficient on the local volumetric water content in a porous material. As assumed above it is considered isothermal absorption in a material which can be assumed homogeneous at the macroscopic scale. It can be assumed that the behavior of water vapour in a homogeneous porous medium should be highly similar to the behavior of water vapour in an individual through-type pore (Fig. \ref{1D-pore_diff}) with specially selected initial and boundary conditions for equation (\ref{Dif1dim}). For further consideration the equation (\ref{Dif1dim}) is used for description water vapour evolution in the pore with the following initial and boundary conditions.
\begin{eqnarray*}
w(t,0)=&w_1&\\
w(t,l_x)=&w_2&  \\
w(0,x)=&\frac{w_2-w_1}{l_x}x+w_1&
\end{eqnarray*}
\vspace{-1cm}
 
\section{Molecular dynamic model}
Molecular dynamic simulation of microscopic system evolution is based on representation of object modeling as a system of interacting particles (atoms or molecules). The evolution of the system is a result of a motion of the particles mentioned above. Particle coordinates in each subsequent time step are calculated by integration of equations of motion. These equations contain potentials of particle interactions with each other and with an external environment.

In classical molecular dynamics, the behavior of an individual particle is described by the Newton equations of motion \cite{Gould}, which can be written in the following form
\begin{equation}\label{a} 
  m_i\frac{d^2 \vec{r_i}}{dt^2}=\vec{f_i},
\end{equation}
\noindent
where $i \ - $ a particle number, $(1\leq i \leq N)$, $N \ - $ the total number of particles, $m_i \ - $ particle mass, $\vec{r_i}\ - $ coordinates of position,  $\vec{f_i} \ - $ the resultant of all forces acting on the particle. This resultant force has the following representation
\begin{equation}\label{b} 
  \vec{f_i} = -\frac{\partial U(\vec{r_1},\ldots,\vec{r_N})}{\partial \vec{r_i}} + \vec{f_i}^{ex},
\end{equation}
where $U \ - $ the potential of particle interaction, $\vec{f_i}^{ex} - $ a force caused by external fields. 

\begin{figure}[H]
\begin{center}
\includegraphics[width=1.0\linewidth]{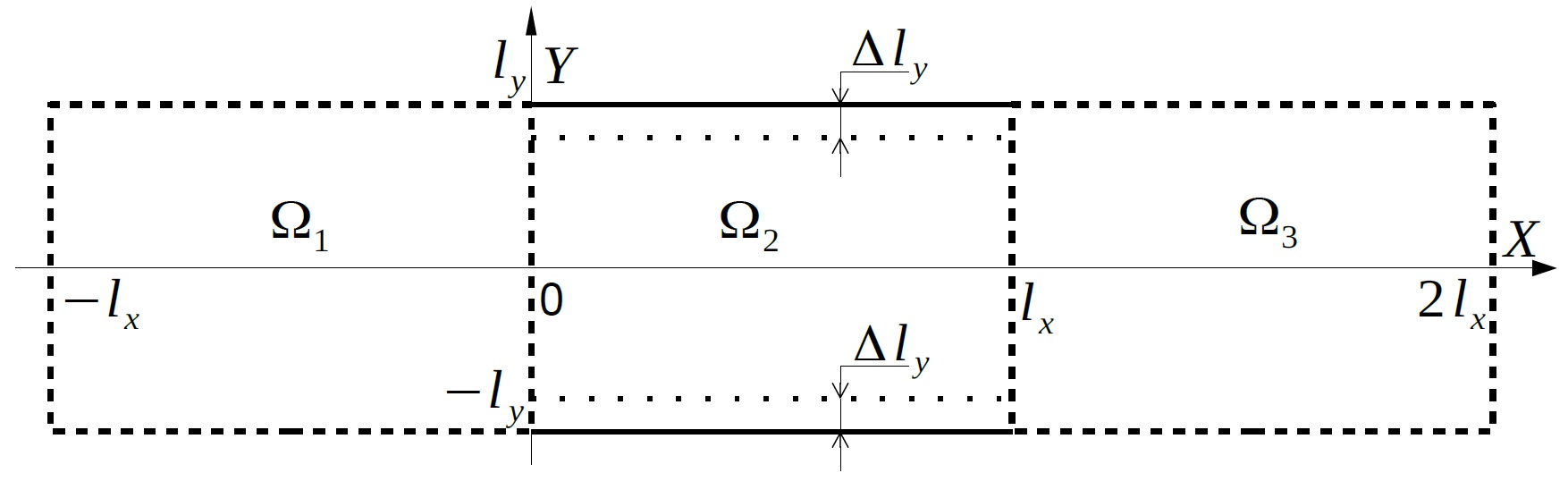}
\end{center}
\vspace{-1cm}
\caption{The geometrical configuration of molecular dynamic model of the through-type pore }
\label{MD-pore_model}
\end{figure}

It is used the following geometrical configuration (Fig. \ref{MD-pore_model}) for molecular dynamic model of water vapour diffusion through the pore.
It takes the following boundary conditions for further molecular dynamic simulations. The periodic boundary conditions have been fixed for boundaries $\Gamma_{11}=[-l_x\leq x \leq 0; -l_y]$, $\Gamma_{12}=[-l_x; -l_y\leq y\leq ly]$, $\Gamma_{13}=[-l_x\leq x \leq 0; l_y]$, $\Gamma_{31}=[l_x\leq x \leq 2l_x; -l_y]$, $\Gamma_{32}=[2l_x; -l_y\leq y\leq ly]$, $\Gamma_{33}=[l_x\leq x \leq 2l_x; l_y]$. There are no boundaries between $\Omega_1$, $\Omega_2$ and $\Omega_3$ domains. $\Gamma_{21}=[0\leq x \leq l_x; -l_y]$, $\Gamma_{22}=[0\leq x \leq l_x; l_y]$ are absolutely non transparent for water molecules boundaries (''walls'') of the pore.

For a simulation of particle interaction in $\Omega_1$, $\Omega_2$ and $\Omega_3$ domains is used the Lennard-Jones potential \cite{LJ}
\begin{equation}
U(r)=4\varepsilon\left[\left(\frac{\sigma}{r}\right)^{12}-\left(\frac{\sigma}{r}\right)^{6}\right]
\end{equation}
with $\sigma = 3.17 \mbox{\AA}$ and $\varepsilon = 6.74\cdot 10^{-3}$ eV. 

Here $r$ -- the distance between the centers of the particles, $\varepsilon$ -- the depth of the potential well, $\sigma$ -- the distance, where the energy of interaction becomes equal to zero. Parameters $\varepsilon$ and $\sigma$ are characteristic for each sort of atom. The minimum of the potential is reached when $r_{min} = \sigma\sqrt[6]{2}$.
It is the most used to describe the evolution of water in liquid and saturated vapour form.

The interaction between particles and the ''walls'' are described by one of the well-known potential (\ref{LJ_9-3}) so called Lennard-Jones 9-3 wall potential \cite{Siderius2011}.
\begin{equation}
U_{9-3}(\xi)=2\pi\rho_w\sigma_{wf}^3\varepsilon_{wf}\left[\frac{2}{45}\left(\frac{\sigma_{wf}}{\xi}\right)^9-\frac{1}{3}\left(\frac{\sigma_{wf}}{\xi}\right)^3\right]
\label{LJ_9-3} 
\end{equation}
Here $\sigma_{wf}$ and $\varepsilon_{wf}$ are the Lennard-Jones length and energy parameters for the interaction of a water molecule with a single constituent particle of the solid, and $\xi$ is the perpendicular distance between the fluid molecule and the material surface. In the case of two-dimensional model of a through-type pore $\xi$ is $\Delta y$, the distance from the water molecule to the pore wall along the $y$ coordinate. $\rho_w$ is the volume density of the solid.

Equations of motion (\ref{a}) were integrated by Velocity Verlet method \cite{Verlet1967}.
Within the framework of the Velocity Verlet method integrating the equations of motion is performed as follows:
\begin{itemize}
\item At the beginning of each step values $r(t),\ v(t),\ f(t) $ at the time $ t $ are defined or calculated in the previous step.

\item First, the coordinates of the particle's new location are calculated at time $t+\Delta t$ , then the velocities of particles  are calculated at time $t+\frac{\Delta t}{2}$
\begin{eqnarray}
r(t+\Delta t) & = & r(t)+\Delta t\  v\left(t\right)+\frac{ (\Delta t)^2}{2}  a(t), \nonumber\\
v\left(t+\frac{\Delta t}{2}\right) & = & v(t)+\frac{\Delta t}{2}\ \frac{f(t)}{m}.\nonumber 
\end{eqnarray}

\item Next, the forces $f(t)$ acting on the particles are recalculated at time $t+\Delta t$.

\item Finally, the values of velocities are calculated at time $t+\Delta t$
$$
v(t+\Delta t) = v\left(t+\frac{\Delta t}{2}\right)+\frac{\Delta t}{2}\ \frac{f(t+\Delta t)}{m}.
$$

\end{itemize}
One of the important thing for the molecular dynamic simulations to account for the effects of energy interchanging with the ambient is a special algorithm which is called a thermostat. In this work Berendsen thermostat \cite{Berendsen1984} is used for temperature calibration and control in all domains. 
This thermostat uses alternating nonlinear friction in the equations of motion and is realized by the following equations.
\begin{eqnarray}
 \frac{dr_{i}(t)}{dt}&=&v_{i}(t), \nonumber\\
 \frac{dv_{i}(t)}{dt}&=&\frac{f_{i}(t)}{m_{i}}-\lambda(t)v_{i}(t).\label{ev2}
\end{eqnarray}
The coefficient of the velocity recalculation $\lambda(t)$ at every time step $t$ 
\begin{equation}
 \lambda(t)=\left[1+ \frac{\Delta t}{\tau_B}\left(\frac{T_{0}}{T(t)}-1\right) \right]^{\frac{1}{2}}.
\end{equation} 
depends on the so called ''rise time'' of the thermostat $\tau_B$ which belongs to the interval $ [0.1,2] \ \mbox{ps}$.

$\tau_B$ describes strength of the coupling of the system to a hypothetical heat bath. For increasing $\tau_B$, the coupling weakens, i.e. it takes longer to achieve given temperature $T_0$ from current temperature $T(t).$

The Berendsen algorithm is simple to implement and it is very efficient for reaching the desired temperature from far-from-equilibrium configurations. 

The Andersen thermostat \cite{Andersen1980} is used for simulation of interaction of water molecules with walls of the pore to account for the sticking effect of vapour particles to pore walls. In addition the Andersen thermostat is the simplest one which does correctly sample the NVT ensemble. This thermostat acts on particles entered into a thin layer of width $\Delta l_y$ near the pore walls. And at each step, some prescribed number of particles is selected, and their momenta (actually, their velocities) are drawn from a Gaussian distribution at the prescribed temperature.

This is intended to imitate collisions with bath particles at a specified $T$. The strength of the coupling to the heat bath is specified by a collision frequency, $\nu$. For each particle, a random variate is selected between $0$ and $1$. If this variate is less than $\nu \Delta t$, where $\Delta t$ is a time step, then that particle's momenta are reset. In this case, the Velocity Verlet integration scheme with the use of the Andersen thermostat can be represented as follows.
\begin{itemize}
\item Calculation of particle position for the time moment $t+\Delta t$.
$$
\mathbf{r}(t+\Delta t)=\mathbf{r}(t)+\mathbf{v}(t)\Delta t +\frac{\mathbf{f}(t)}{2m}(\Delta t)^2
$$
\item Calculation of particle velocity for the time moment $t+\frac{\Delta t}{2}$.
$$
\mathbf{v}(t+\frac{\Delta t}{2})=\mathbf{v}(t)+\frac{\mathbf{f}(t)}{2m}\Delta t
$$
\item And the final calculation of particle velocity for the time moment $t+\Delta t$.
$$
\mathbf{v}(t+\Delta t)=\mathbf{v}(t+\frac{\Delta t}{2})+\frac{\mathbf{f}(t+\Delta t)}{2m}\Delta t
$$
\item Particle velocity reset.

If $\xi < \nu\Delta t$, where $\xi \in U[0;1]$ and $U[0;1]$ numerical uniform distribution, then the value of the velocity $\mathbf{v}(t+\Delta t)$ is set to the random number $\zeta\in N[1;T]$, where $N[1;T]$ is Gaussian distribution with variance equal $T$.
\end{itemize}

The pressure in the pore was controlled using the formula based on virial equation \cite{Frenkel2002}.
$$
P = \frac{1}{3V}\left(\left\langle 2K\right\rangle-\left\langle\sum\limits_{i<j} r_{ij}\cdot f\left(r_{ij}\right)\right\rangle\right).
$$
Here $V$ is the pore volume, $\left\langle 2K\right\rangle$ is the doubled kinetic energy averaged over the ensemble, $f\left(r_{ij}\right)$ is the force between  particles $i$ and $j$ at a distance $r_{ij}$.
 
\section{Computer simulations}
In this paper it is used a flat pore model with dimensions of $1\mu m \times 1\mu m \times d$, where $d$ is the thickness of one layer of water molecules equal to the diameter of the molecule. Since it is assumed that the motion of molecules along the $z$ axis does not occur, this model is actually two-dimensional. The concentration of molecules, thus, is calculated as three-dimensional one, considering that water molecules are located in the volume $V = 1\mu m \times 1\mu m \times d$. Everywhere in this article it is used reduced units which is marked by $^{*}$. 

If we consider that water molecules are located in the volume $V^{*} = 3160 \times 3160 \times 1$ then initial concentrations were obtained from the density of water vapour at the appropriate pressure and density at a given temperature using known tabulated data. Specifically for temperature  $T_{0}=35^{o}C$ $(T_{0}^{*}=3.94)$ density of saturated water vapour is 0.03962 $kg/m^{3}$. In this case  the  pore  contains 420 water molecules which are evenly distributed in the pore at the initial time moment. Initial velocities are randomly selected from $ [0,1] $,  then the velocities are set so that the total velocity (momentum) is equal to zero, and then they are changed to achieve the desired initial temperature $T_{0}$.

We consider that on the left side of the pore (Fig. \ref{1D-pore_diff}) we have 100 \% saturated water vapour for temperature $35^{o}C$. It means that the pressure $p_L = 5.622\, kPa$ and density is equal to $0.03962\, kg/m^{3}$. On the right side of the pore  temperature is the same, but pressure is lower $p_R = 2.337\, kPa$ that means  that there is less number of water molecules (there is not 100 \% saturated water vapour) in the right part of the pore. In reduced units we have $p_L^{*} = 1.65183\cdot 10^{-4}$, $p_R^{*} = 0.686647\cdot 10^{-4}$. We assume that pressure changes his value linear from the highest to the lowest value inside the pore	

$$
p = \left(1.65183 - \frac{0.965183}{3160}x\right)\cdot 10^{-4}, 0\leq x \leq 3160.
$$			

The difference between the pressure values on the left and right ends of the pore and on the right leads to the appearance of external force acting on each particle in the pore. This force is also a linear function of the variable $x$. It  can be considered as the second part of the sum (\ref{b}). In general we can compute this force as  $F = p\cdot S$, where $p$ is pressure and $S$ is the surface to which the pressure is applied. In our example, the surface is a cross-section of the water molecule $S=\pi(d/2)^{2}$ or $ S^{*}=\pi/4$ and external force is

$$
F = \frac{\pi}{4} \left(1.65183 - \frac{0.965183}{3160}x\right)\cdot 10^{-4} .
$$

The conclusion is that every particle in the pore is subject to the force from interaction  with all other particles in the pore by Lennard-Jones potential  and the force interaction with walls of pore by Lennard-Jones 9-3 potential  and external force caused by difference of pressures.

On the left and the right side of pore  (Fig. \ref{1D-pore_diff}) are implement periodic boundary conditions and up and down side of pore is from solid material from silicon with $\sigma=0.420\, nm$ and $\varepsilon=32.5818\cdot 10^{-23} J$, thus $\sigma_w^{*}=1.327 $ and $\varepsilon_w^{*}=0.3018$. We used Lorentz-Berthelot rules  \cite{Nezbeda2003} to compute parameters for Lennard-Jones 9-3 potential for interaction particles and wall,$\sigma_{wf}^{*}=1.1635 $ and $\varepsilon_{wf}^{*}=0.5494$.  Density of pore wall is $\rho_w=2330\, kg/m^{3}$,   $\rho_w^{*}=21.00541$.

Our run has $2000000$ time steps with length $dt^{*} = 0.005$, we reach result time $16663\, ps$. The Berendsen thermostat works with parameter $\tau_B=0.5\, ps$. The Andersen thermostat uses $\nu=1$ and it works in layer $\Delta l_y=2\sigma$.

\begin{figure}[H]
\begin{center}
\includegraphics[width=0.6\linewidth]{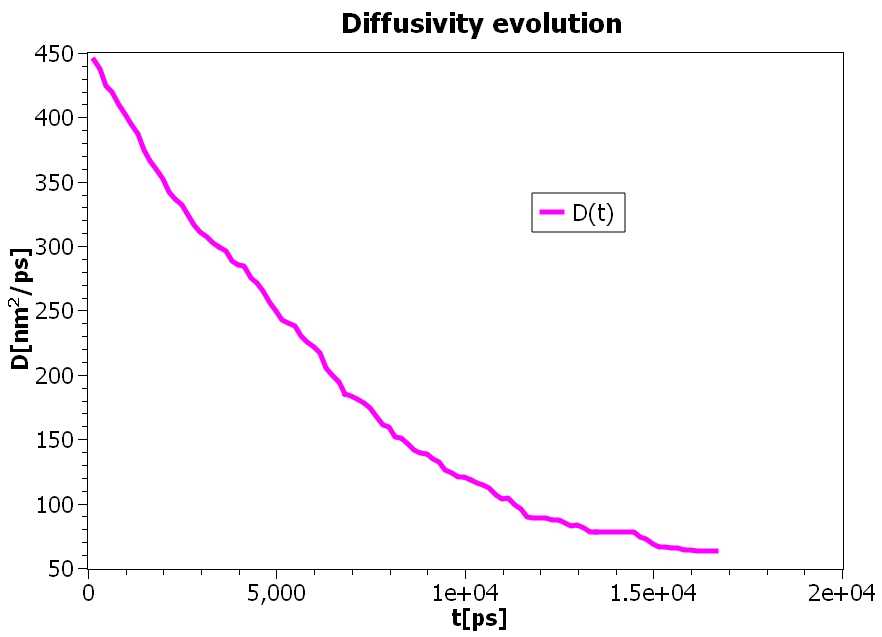}
\end{center}
\caption{Time evolution of averaged over pore volume diffusivity}
\label{DiffEvol}
\end{figure}

Time evolution of averaged over pore volume diffusivity is presented in Figure \ref{DiffEvol}.

\begin{figure}[H]
\begin{center}
\includegraphics[width=0.6\linewidth]{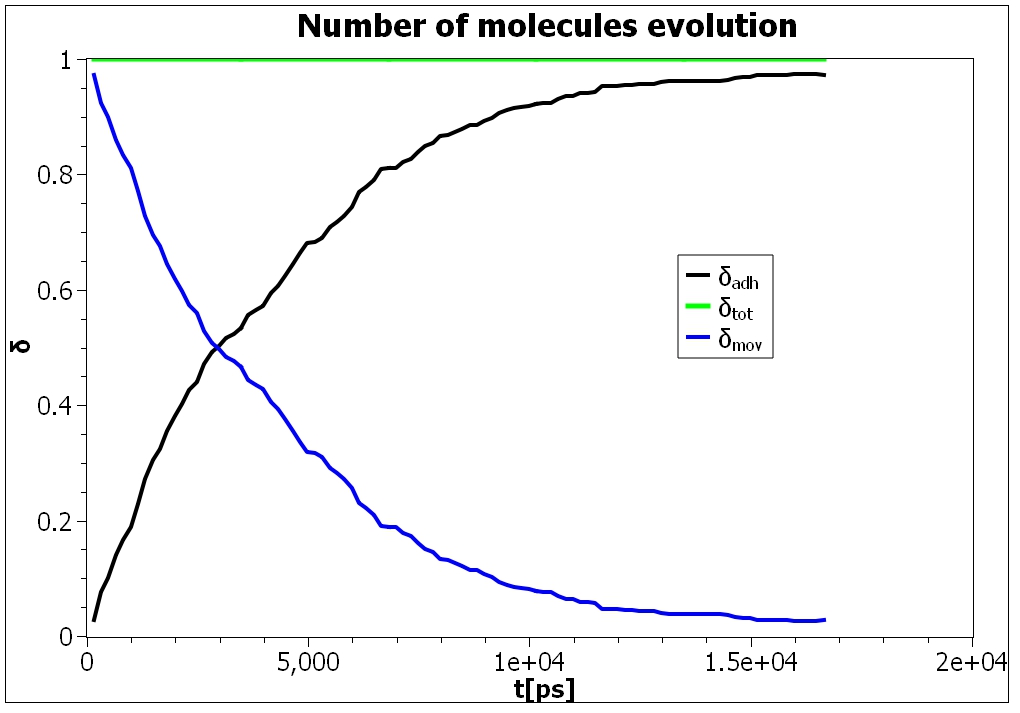}
\end{center}
\caption{Time evolution of the ratio of the number of stuck and moving molecules to the total number of molecules in the pore. $\delta_{adh}=N_{adh}/N_{tot}$, $\delta_{mov}=N_{mov}/N_{tot}$, $N_{adh}$ - a number of water molecules stuck to the pore walls, $N_{mov}$ - a number of moving water molecules in the pore, $\delta_{tot}=N_{tot}/N_{tot}=1$.}
\label{NumMolEvol}
\end{figure}

Time evolution of the ratio of the number of stuck and moving molecules to the total number of molecules in the pore is presented in Figure \ref{NumMolEvol}.

\begin{figure}[H]
\begin{center}
\includegraphics[width=0.6\linewidth]{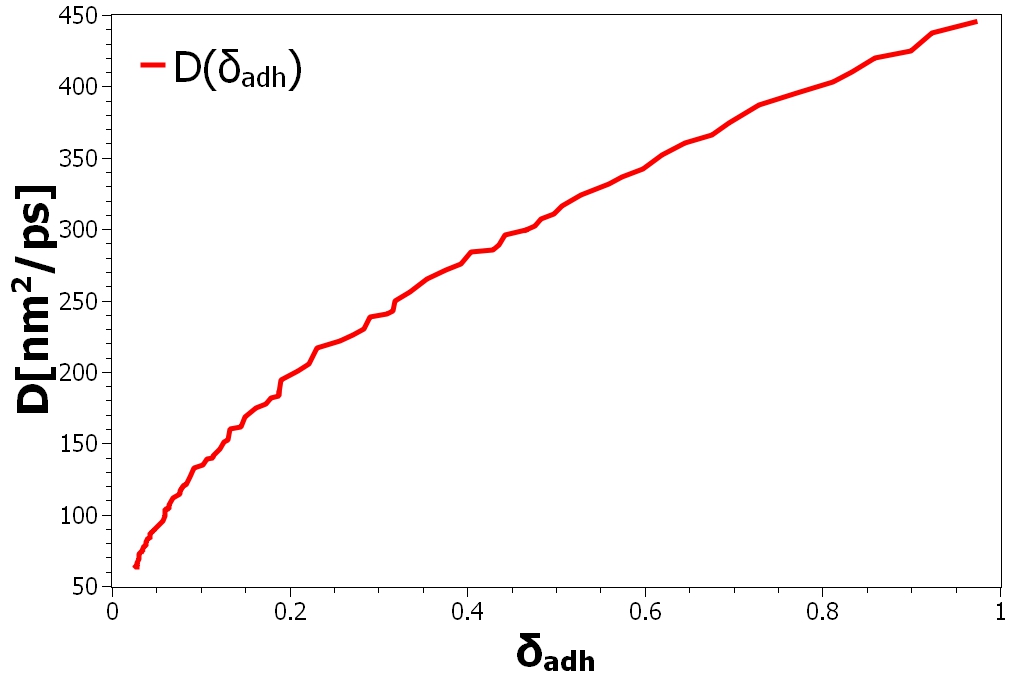}
\end{center}
\caption{Dependence of the average diffusivity on the ratio of the number of adherent molecules to the total number of molecules}
\label{DiffAdh}
\end{figure}

\begin{figure}[H]
\begin{center}
\includegraphics[width=0.6\linewidth]{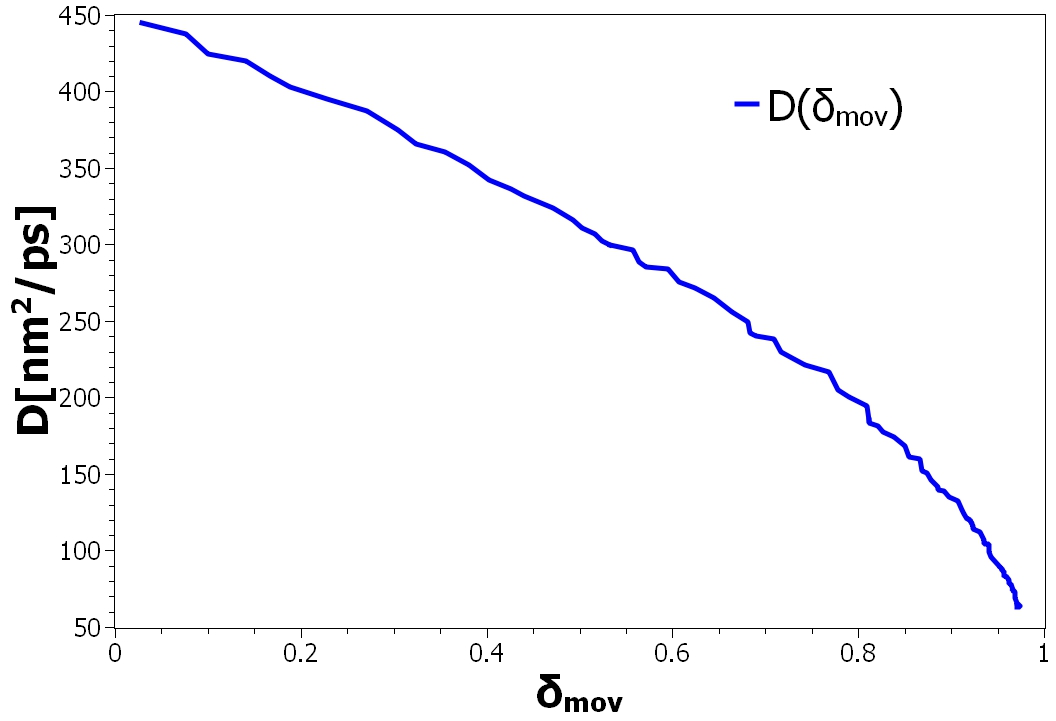}
\end{center}
\caption{Dependence of the average diffusivity on the ratio of the number of non adherent molecules to the total number of molecules}
\label{DiffMov}
\end{figure}

Dependence of the average diffusivity on the ratio of the number of stuck and non stuck  molecules to the total number of water molecules are presented in Figures \ref{DiffAdh} and \ref{DiffMov} correspondingly.

\begin{figure}[H]
\begin{center}
\includegraphics[width=1.0\linewidth]{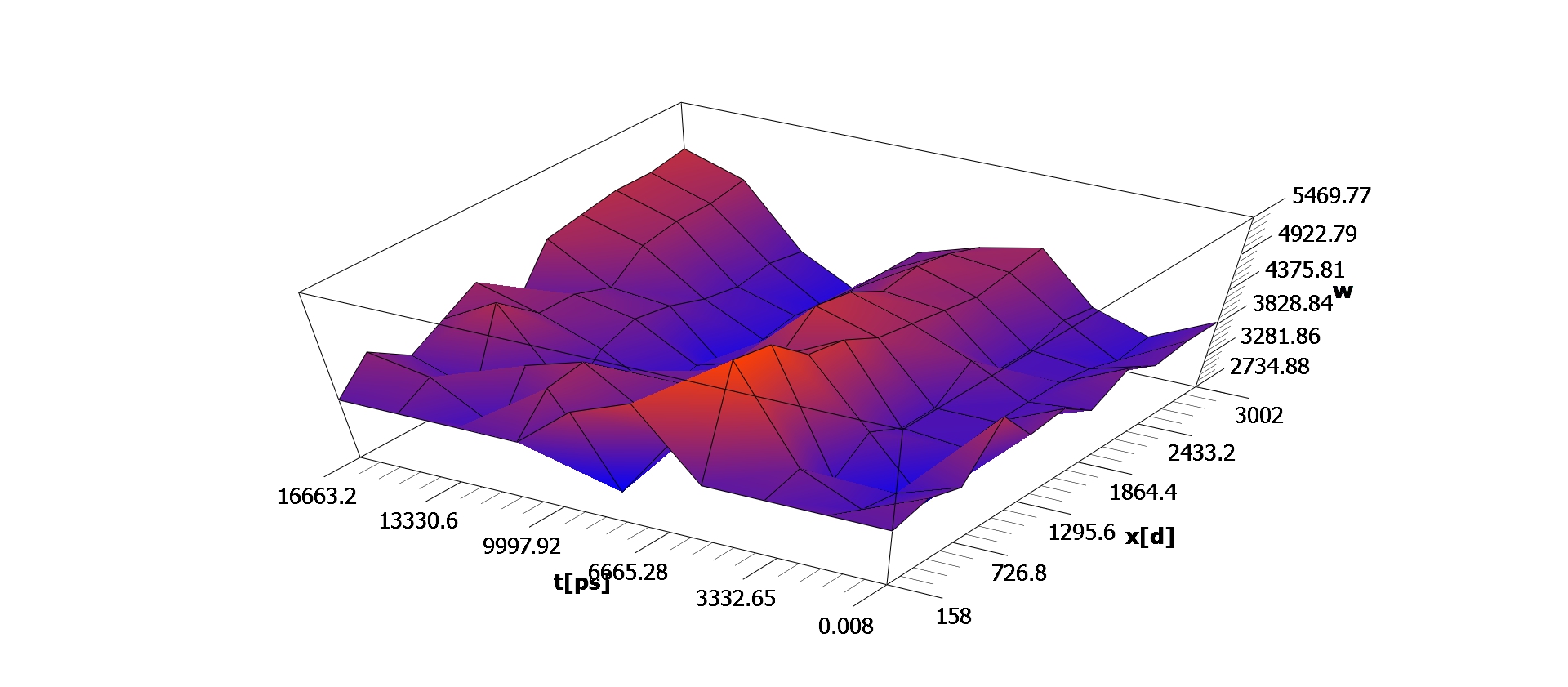}
\end{center}
\caption{Dependence of the local density $w[ng/nm^3\times 10^{-14)}]$ of water molecules on $x$-coordinate along the pore and a time. $d$ - a diameter of water molecule.}
\label{Density_surf.jpg}
\end{figure}

The two-dimensional density profile $D(x,t)$ is presented in Figure \ref{Density_surf.jpg}.

\section{Conclusions}
As follows from the simulation results behavior of wet water vapour in the through-type pore is very similar to moisture behavior in homogeneous porous material. Diffusivity is almost exponential time function (Fig. \ref{DiffEvol}) and also non linearly, up to model discretization, depends on the $\delta_{adh}=N_{adh}/N_{tot}$, the ratio of stuck water molecules number to total number of water molecules in the pore (Fig. \ref{DiffAdh}). It is very similar to dependency of the diffusivity on local volumetric water content inside the pore. Further improvement of the model should lead to more realistic description of wet water vapour behavior inside the through-type pore and will be able to more accurately describe the experimental data for isothermal absorption in a material which can be assumed homogeneous at the macroscopic scale and where gravity is negligible when compared to the capillary forces.

\end{document}